# Contribution à la modélisation explicite des plates-formes d'exécution pour l'IDM

**Frédéric Thomas\* — Jérôme Delatour\*\* — Francois Terrier\* — Matthias Brun\*\* — Sébastien Gérard\***

*\* CEA LIST*
*Boîte courrier 94 - F-91191 Gif sur Yvette Cedex*
*{prenom.nom}@cea.fr*

*\*\* ESEO*
*4 rue Merlet de la Boulaye, BP 30926, F-49009 ANGERS cedex 01*
*{prenom.nom}@eseo.fr*

RÉSUMÉ. *L'un des principes de l'Ingénierie Dirigée par les Modèles (IDM) est de mettre en œuvre un même modèle de fonctionnalités sur différentes solutions technologiques (c.a.d. plate-forme). Certaines de ces plates-formes permettent l'exécution du système. L'un des objectifs de l'IDM est ainsi de pouvoir automatiser les transformations d'une même description indépendante des plates-formes d'exécution vers des descriptions dédiées à des plates-formes particulières. A notre connaissance, peu de travaux se sont à ce jour intéressés à la modélisation des plates-formes d'exécution. Pourtant, une modélisation explicite de ces plateformes favorise la prise en compte de leurs caractéristiques (par ex., performances, maintenance et portage). Cet article recense les travaux menés sur ce sujet et propose un motif pour la description des plateformes d'exécution. Il vise à montrer la faisabilité et l'intérêt d'un métamodèle de plate-forme d'exécution.*

ABSTRACT. *One foundation of the model driven engineering (MDE) is to separate the modelling application description from its technological implementation (i.e. platform). Some of them are dedicated to the system execution. Hence, one promise solution of the MDE is to automate transformations from platform independent models to platform specific models. Little work has explicitly described platform characteristics. Yet, an explicit modelling allows taking in account their characteristics more easily (par ex., performances, maintainability, portability). This paper presents both an execution platform modelling state of art and a pattern to describe execution platform modelling framework. It intends to confirm the feasibility and the interests in describing an execution platform metamodel.*

MOTS-CLÉS : *Ingénierie dirigée par les modèles, plate-forme d'exécution, SRM*

## 1. Introduction

La volonté de s'abstraire des technologies matérielles et logicielles permettant l'exécution d'une application est un souci constant dans l'industrie informatique. La monté en abstraction a prouvé que le développement des systèmes gagnait en productivité et en robustesse. Il permet, par exemple, d'envisager des systèmes réutilisables et plus facilement maintenables. Ainsi, ce besoin a suscité le développement de langages de programmation ayant des pouvoirs d'expression de plus en plus élevés (par ex., les langages procéduraux, les langages objets et les langages de modélisation) et l'utilisation de couches logicielles de plus en plus abstraites (par ex., les systèmes d'exploitation, les machines virtuelles et les intergiciels). Ces couches sont des plates-formes sur lesquelles le concepteur s'appuie à un moment donné du cycle de vie du logiciel. Elles favorisent alors le déploiement ou le portage des applications sur différentes plates-formes d'exécution sous-jacentes. La machine virtuelle JAVA permet par exemple d'exécuter la même description de l'application sur des systèmes d'exploitation différents. De même, la technologie des systèmes d'exploitation permet la mise en œuvre d'une même description logicielle sur de multiples architectures matérielles. Elle facilite ainsi la réutilisation d'application logicielle dont le cycle de vie est souvent plus long que celui des plates-formes d'exécution matérielles. Certaines de ces plates-formes logicielles suivent des normes. Les standards dédiés aux systèmes d'exploitation temps réel tel que POSIX (The Open Group Base Specification, 2004), OSEK/VDX (OSEK/VDX Group, 2005) ou encore ARINC-653 (Airlines electronic engineering comittee, 2003) en sont des exemples. Ces normes ont permis d'unifier leurs interfaces de programmation et donc d'apporter une certaine indépendance vis-à-vis d'un ensemble d'implantation.

Néanmoins, ces solutions ont trouvé leurs limites. D'une part, certains domaines ne permettent pas l'utilisation de couches logicielles supplémentaires. C'est souvent le cas des systèmes embarqués temps réel où l'on apporte un grand intérêt aux performances du système qui englobe à la fois la plate-forme d'exécution et l'application (c.a.d. la description des fonctionnalités). Ces empilements de couche impliquent l'utilisation de supports d'exécution de plus grandes capacités (par ex., mémoire, processeur et consommation d'énergie) et limitent la prédictibilité (en particulier du point de vue temporel) du futur système. D'autre part, le concept d'une interface de programmation unique et universelle est une utopie. Dans la plupart des cas, les interfaces standardisées ne sont implantées que jusqu'à un certain degré de conformité. De plus, les normes définissent par essence des points de variation sémantique qui impliquent des comportements d'exécution différents. La prise en compte des plates-formes d'exécution dans le cycle de développement d'un système reste donc bien souvent du cas par cas, ce qui limite l'automatisation des déploiements.

Les travaux liés à l'ingénierie dirigée par les modèles (IDM) tentent d'apporter une alternative à ces solutions. Par essence, ces travaux sont basés sur les mêmes idées fondamentales: augmenter le niveau d'abstraction et favoriser l'automatisation. Une des solutions promises est la séparation des préoccupations de modélisation de l'application vis-à-vis du déploiement sur la ou les plates-formes cibles. Ils visent ainsi à proposer un cadre méthodologique et technologique pour faciliter la prise en compte progressive des différentes plates-formes tout en conservant un paradigme unificateur : le modèle (c.a.d. l'abstraction) et les transformations de modèle (c.a.d. l'automatisation). Certains modèles sont indépendants d'un ensemble de plates-formes. Les transformations les rendent ensuite spécifiques à un ensemble de plates-formes.

De nombreux travaux ont été menés pour utiliser l'ingénierie dirigée par les modèles dans le cycle de vie du logiciel. Une approche commune est la description d'une transformation par plate-forme d'exécution cible. Ainsi, la description d'une transformation reste souvent un amalgame de caractéristiques inhérentes à la plate-forme d'exécution et aux technologies de transformations (par ex., transformations de graphes, descriptions de règles de transformation impératives et/ou déclaratives). Les transformations sont alors figées autour d'une plate-forme cible et d'un ensemble de règles métiers connues lors de leurs descriptions. Leurs mises en œuvre, leurs maintenances et leurs optimisations sont alors réservées à des experts à la fois de l'IDM, du domaine métier (par ex., du temps réel embarqué) et des plates-formes d'exécution envisagées. Les interactions avec ces transformations pour d'éventuelles adaptations ou optimisations restent laborieuses et rendent difficile le processus d'automatisation.

En théorie, une modélisation explicite favorise la prise en compte de leurs caractéristiques. Il permet d'adapter les transformations aux besoins des utilisateurs, d'automatiser la conception d'application exécutable, d'en maitriser le comportement, d'en vérifier le fonctionnement, d'en faciliter le portage tout en conservant des transformations génériques. Toutefois, peu de travaux ont porté sur de tels modèles. Nous proposons de contribuer à la description d'un métamodèle de plate-forme d'exécution.

Nous nous sommes appuyés sur différentes expérimentations de transformations de modèles (Thomas et *al.*, 2007) (Thomas et *al.*, 2006) (Delatour et *al.*, 2005). Ces expérimentations visent à expliciter un modèle de plate-forme d'exécution pour des technologies dédiées à l'embarqué temps réel (c.a.d. système d'exploitation temps réel embarqué). Le métamodèle de plate-forme d'exécution résultant est intégré dans la définition du profile UML MARTE (OMG, 2007). Il permet la modélisation des plates-formes d'exécution logicielles embarquées temps réel.

Dans une première partie nous identifions ce qu'est une plate-forme d'exécution au sens de l'IDM. Dans une seconde partie, nous identifions les principales caractéristiques de ces plates-formes en nous appuyant sur les travaux de la communauté. Dans une troisième partie nous illustrons d'une part, l'implantation d'un métamodèle pour décrire des plates-formes d'exécution et d'autre part, nous décrivons une première expérimentation sur la description d'une transformation manipulant des modèles explicites de plate-forme d'exécution. Enfin, nous discutons de nos retours d'expérience sur la modélisation explicite de la plate-forme d'exécution avant de dresser les perspectives de ces travaux.

## 2. Qu'est ce qu'une plate-forme d'exécution?

Il existe aujourd'hui un grand nombre d'entités identifiées comme étant des plates-formes d'exécution. Les définitions sont souvent dédiées au domaine d'application. Nous présentons dans un premier temps ce qu'est une plate-forme d'exécution dans l'industrie. Dans un second temps nous nous intéresserons à définir le concept de plate-forme d'exécution au sens de l'IDM.

### 2.1. Le concept de plate-forme d'exécution dans l'industrie

Historiquement dans l'industrie, le concept de plate-forme d'exécution réfère essentiellement à la structure matérielle qui supporte l'exécution du logiciel. L'apparition du style architectural en couches a étendu l'usage de ce concept pour décrire de la même façon des supports logiciels. La figure 1 illustre une proposition de C. Atkinson faite dans (Atkinson *et al.*, 2005) et décrivant une hiérarchie typique des plates-formes d'exécution : les infrastructures matérielles, les systèmes d'exploitation, les machines virtuelles et les intergiciels.

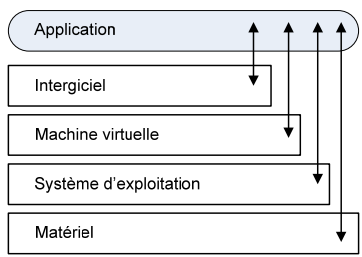

**Figure 1** *Hiérarchie typique de plates-formes d'exécution*

De manière générale dans l'industrie, une plate-forme est donc une infrastructure, un système, permettant l'exécution d'un autre système appelé application.

En pratique, la distinction entre plate-forme et application dépend du point de vue où l'on se place. Toute plate-forme peut être considérée comme une application pour une plate-forme de plus bas niveau d'abstraction (Marvie *et al.*, 2006). Ainsi, un fournisseur de plate-forme matérielle considère le système d'exploitation comme une application s'exécutant sur la plate-forme matérielle. En revanche, un développeur logiciel perçoit un système d'exploitation comme la plate-forme (logicielle) exécutant son application. La caractérisation d'un système comme étant une plate-forme n'est donc pas liée à sa nature, mais à son rôle dans le système global.

## 2.2. Plate-forme et IDM

L'OMG (Object Management Group) est un consortium international de normalisation très actif au sein de la communauté de l'IDM. Il a notamment normalisé le langage de modélisation unifié UML (« Unified Modeling Language », (OMG, 2007)), mais aussi définit un cadre normatif pour l'IDM : le « Model Driven Architecture » (Architecture Dirigée par les Modèles, MDA, (OMG, 2003)). L'une des promesses du MDA est de pouvoir déployer des applications logicielles facilement sur des plates-formes d'exécution différentes. A partir de descriptions indépendantes des plates-formes (PIM : « Platform Independent Model »), elle vise à générer des applicatifs dédiés (PSM : « Platform Specific Model ») à une plate-forme technologique donnée. Le MDA permet ainsi en théorie de mettre en œuvre un même modèle de fonctionnalités (également appelé modèle métier) sur différentes solutions technologiques (c.a.d. des plates-formes d'exécution différentes).

Le MDA est aujourd'hui une branche particulière de l'IDM. Néanmoins, les définitions qu'il propose, unifient les concepts utilisés par la communauté. Une plate-forme est définie dans (OMG, 2003) comme *« un ensemble de sous-systèmes et de technologies qui fournissent un ensemble cohérent de fonctionnalités au travers d'interfaces et la spécification de patrons d'utilisation que chaque sous-système qui dépend de la plate-forme peut utiliser sans être concerné par les détails et sur la façon dont les fonctionnalités sont implantées ».*

Dans cette définition, les technologies et les sous-systèmes caractérisés comme étant des plates-formes englobent aussi bien l'environnement de développement que l'environnement d'exécution de l'application. Ces deux environnements sont nécessaires pour l'exécution du système, pour l'expression, la représentation, la vérification de propriétés et même pour la création de modèle (Atkinson *et al.*, 2005). Les infrastructures d'exécution sont donc des cas particuliers de systèmes nommés plates-formes au sens de l'IDM. Comme toutes plates-formes, les infrastructures d'exécution sont donc des abstractions vis-à-vis des plates-formes sous-jacentes (Sangiovanni-Vincentelli *et al.*, 2004).

Orthogonalement, J.P Almeida (Almeida, 2006) précise que parmi ces plates-formes, certaines sont dites abstraites et d'autres concrètes. Cette notion de plates-formes abstraites fait référence au concept d'éléments abstraits des langages objets. Dans l'approche proposée par J.P Almeida, une plate-forme abstraite est une plate-forme fictive, idéale face aux besoins à un instant donné du développement de l'application. C'est une spécification d'un ensemble de technologies qui permettent une exécution. Une plate-forme concrète est une implémentation d'un ensemble de technologies permettant l'exécution. Un exemple de plate-forme d'exécution abstraite est le langage de description d'application temps réel que nous avons proposé (Concept of Real Time Application CORTA) dans (Delatour *et al.*, 2005). CORTA proposait un ensemble de mécanismes et de services permettant de spécifier des exécutions temps réel. Les concepts manipulés par ce langage étaient donc ceux que l'on retrouve dans la plupart des systèmes d'exploitation temps réel (par ex., tâche et sémaphore). CORTA jouait alors le rôle de langage pivot pour cibler de multiples systèmes d'exploitation temps réel (c.a.d. des plates-formes d'exécution temps réel embarquées) qui eux permettaient une réelle exécution de l'application spécifiée en CORTA.

Indépendamment qu'elles soient abstraites ou concrètes, les plates-formes d'exécution sont au sens de la définition du MDA des serveurs qui par l'intermédiaire de mécanismes offrent des services à d'autres systèmes clients appelés applications. Tant que ces services ne sont pas fournis, il n'est pas possible de spécifier une application exécutable (Selic, 2005). Un modèle de plate-forme d'exécution est donc un modèle de ce serveur, c.a.d, un modèle des mécanismes et des services offerts par la plate-forme.

Contrairement à la modélisation d'une application (c.a.d. d'un client), un modèle de l'implantation des mécanismes et des services d'une plate-forme n'est pas nécessaire. La modélisation de leurs signatures et de leurs comportements visibles depuis l'interface de programmation des applications (« Application Programming Interface », API) suffit. Au sens de l'IDM, un modèle d'une plate-forme d'exécution est donc une représentation de l'API de la plate-forme.

## 3. Un métamodèle pour les plates-formes d'exécution

La modélisation d'une plate-forme d'exécution passe donc par la caractérisation de son API. Dans une première partie, nous référençons les principaux travaux dédiés à la modélisation d'API de plate-forme d'exécution. Nous en déduisons ensuite les principales exigences pour la conception d'un métamodèle de plate-forme d'exécution.

### 3.1. Etat de l'art

C. Atkinson dans (Atkinson *et al.*, 2005) propose de caractériser les plates-formes par :

- le langage de programmation avec lequel est défini l'interface de programmation ;
- les types prédéfinis fournis par la plate-forme. En effet, la plupart des langages informatiques utilisés pour décrire les interfaces des plates-formes d'exécution sont typés. Certains concepts proposés par les plates-formes apparaissent donc comme des types prédéfinis qu'il est possible d'instancier pour mettre en œuvre les fonctionnalités particulières de la plate-forme. Par exemple, un système d'exploitation multitâche propose le concept de tâche. Les tâches sont matérialisées par des types instanciés par l'application. La spécification POSIX propose par exemple le type *pthread_t* pour la manipulation des tâches logicielles ;

- les instances prédéfinies fournies par la plate-forme. D'autres concepts n'apparaissent pas comme des types mais comme des éléments typés. Dans le cadre d'une interface décrite avec un langage orienté objet, ces objets apparaissent comme des instances. Les réveils prédéfinis de certains systèmes d'exploitation et les objets d'entrée/sortie (in, out et err) de la librairie des flux de données en JAVA sont des exemples d'instance prédéfinies ;

- les patrons d'utilisation qui permettent de décrire les règles de bonne utilisation pour manipuler les concepts et les services de la plate-forme. Par exemple, les systèmes d'exploitation temps réel listent souvent un certain nombre de services bloquants qu'il est interdit d'appeler depuis une routine d'interruption.

La figure 2 illustre les liens que fait C. Atkinson entre ces quatre vues.

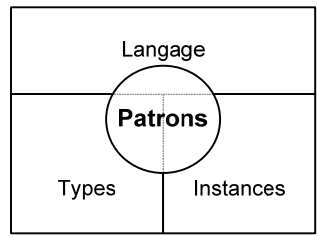

**Figure 2.** *Modèle de caractérisation des plates-formes d'exécution selon C. Atkinson (extrait de Atkinson et al, 2005 )*

Cette vision cible la caractérisation des interfaces de programmation orientées objets. De plus, il n'y a par exemple aucun élément dans cette approche sur la modélisation des traitements qu'offre la plate-forme.

Dans (Marvie *et al.*, 2006), R. Marvie a proposé trois vues gros grains pour modéliser les plates-formes d'exécution :

- la caractérisation des concepts fournis. Un système d'exploitation propose les concepts de fichier et de page mémoire par exemple ;

- la caractérisation des traitements (c.a.d. les services) réalisés par la plate-forme (par ex., la gestion des interactions, la gestion des traitements ou la gestion de la cohérence des données). Par exemple, un système d'exploitation offre des services de lecture et d'écriture dans des fichiers et des pages mémoires ;

- la caractérisation du cycle de vie des concepts et des traitements. Par exemple, le cycle de vie d'un fichier est borné par sa création et sa destruction.

Dans cette approche, nous retrouvons la volonté de caractériser la signature structurelle et le comportement observable (d'un point de vue utilisateur) des plates-formes d'exécution. Cela passe par la modélisation à la fois des concepts et des traitements offerts par la plate-forme.

B. Selic (Selic, 2000) (Selic, 2005) et plus globalement l'OMG caractérisent les concepts et les traitements respectivement comme des ressources et des services. Dans le cadre des profils UML standard (OMG, 2001) puis (OMG, 2007) dédiés à la conception des applications temps réel, l'OMG propose un modèle de ressource particulier (figure 3). Les ressources sont instanciables et offrent des services. Les ressources, les instances de ressources et les services peuvent être qualifiés par des propriétés dites non-fonctionnelles (NFP). Ces NFP traduisent des caractéristiques de performance et de qualité de service qui permettent de qualifier les performances de la plate-forme considérée et de guider le déploiement de l'application sur la plate-forme.

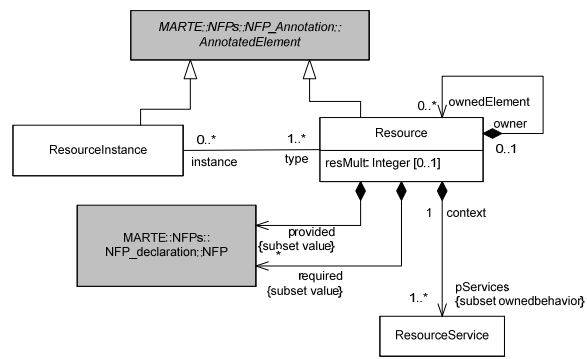

**Figure 3.** *Modèle de ressource du profile MARTE (extrait de (OMG, 2007))*

Nous pensons que toutes ces approches sont complémentaires, ce qui nous permet d'envisager un métamodèle permettant de décrire des plates-formes d'exécution. Nous proposons d'en synthétiser les grandes caractéristiques.

### 3.2. Les contours d'un métamodèle des plates-formes d'exécution

#### 3.2.1 La caractérisation des concepts

Nous nous intéressons ici à la caractérisation des mécanismes fournis par la plate-forme. Nous proposons donc de considérer les types et éléments typés respectivement comme des *Resource* et des *ResourceInstance*. En effet, ce sont des entités que la plate-forme offre en nombre limité et qu'il faut savoir gérer. Certaines sont des types de ressources que l'utilisateur de la plate-forme doit instancier, d'autres sont d'ores et déjà typées, instanciées, par la plate-forme d'exécution. Ces ressources possèdent des propriétés nécessaires pour paramétrer l'exécution (par ex., la priorité d'un traitement et la période d'un traitement périodique).

### 3.2.2 La caractérisation des traitements

Dans la description d'une plate-forme, les traitements seront caractérisés au travers du concept de service. Ainsi, chaque ressource d'une plate-forme offrira au minimum un service. Par exemple, une ressource de type *Fichier* proposera, des services permettant la lecture et l'écriture dans un fichier. Les services peuvent être qualifiés, annotés, par des propriétés non-fonctionnelles (c.a.d. des propriétés qui exprime une qualité de service garantit comme par exemple la latence et le pire temps d'exécution). Ces propriétés reflètent alors les caractéristiques d'exécution et de performances offertes par la plate-forme.

### 3.2.3 La caractérisation des règles d'utilisation de la plate-forme

Une description de la plate-forme consiste non-seulement à décrire les ressources et services offerts par la plate-forme mais aussi la manière d'utilisation de ces concepts. Les règles d'utilisation d'une plate-forme sont aussi bien des contraintes que des guides de bonne utilisation. Ces règles sont internes à la plate-forme. Elles sont souvent décrites en langage naturel sous forme de contraintes dans le manuel d'utilisation. Elles peuvent être caractérisées par un langage de contrainte tel que l'Object Constraint Language (OMG, 2006).

### 3.2.4 La caractérisation du cycle de vie des ressources et services

La caractérisation du cycle de vie des ressources et services passe par la description d'un comportement observable des ressources et des services depuis l'interface de programmation de la plate-forme. Ces descriptions comportementales sont souvent associées à des modèles de calcul qui spécifient le modèle d'exécution supporté par la plate-forme. Dans le contexte d'une description basée UML, des modèles d'activités et/ou des modèles d'états-transition peuvent contribuer à leurs modélisations.

### 3.2.5 L'ouverture sur différents styles de modélisation

Outre la proposition d'artéfacts adéquates pour décrire les quatre grands axes précédents, un métamodèle de plate-forme ne doit pas imposer un style de modélisation particulier (par ex., une modélisation fonctionnelle, une modélisation orientée objet ou enfin une modélisation orientée composant). Les technologies d'implémentation des plates-formes d'exécution étant hétérogènes (procédurale, objet et composant), les formalismes adéquates pour décrire les API de plates-formes le sont tout autant. Le métamodèle de plate-forme d'exécution doit donc permettre l'expression de cette hétérogénéité.

## 4. Un métamodèle détaillé de plate-forme d'exécution temps réel embarqué

Le métamodèle résultant des points précédents peut être implanté sous la forme d'un langage spécifique ou bien comme une extension d'UML (c.a.d. un profil

UML). Dans le cadre du consortium ProMARTE[1], visant la description d'un nouveau profil UML dédié à la modélisation et à l'analyse des systèmes temps-réel embarqués (MARTE : « UML Profile for Modeling and Analysis of Real-Time and Embedded systems ») (OMG, 2007), nous nous sommes concentrés sur la définition d'un métamodèle pour la description des ressources et des services des plates-formes d'exécution multitâches temps réel embarquées. Ce métamodèle est traduit sous la forme d'un profil UML nommé « Software Resource Modeling » (SRM).

Un profil est une extension au métamodèle auquel il fait référence, ici UML. Les éléments profilés du métamodèle sont spécialisés par des stéréotypes. Cette technique d'extension (dite extension légère) permet d'adapter la syntaxe et la sémantique d'un langage donné (par ex., UML) à un domaine d'application particulier. Les stéréotypes peuvent posséder des propriétés (également souvent appelées « tag[2] »). Enfin, un profil peut préciser de nouvelles contraintes sur un élément du langage spécialisé pour en limiter/guider l'utilisation.

SRM est limité à la caractérisation des concepts et des traitements. Il permet typiquement de décrire les ressources et services offerts par les systèmes d'exploitation temps-réel multitâches. Ce n'est pas la spécification d'une nouvelle interface idéale de programmation d'applications multitâches (« Application Programing Interface » (API)). De même, il ne définit pas d'artéfacts pour la description des règles d'utilisation du système et pour la modélisation du cycle de vie des ressources et des services. Les concepts natifs d'UML que sont les contraintes peuvent permettre d'exprimer dans un premier temps la modélisation de ces règles d'utilisation. Les activités, les états et les signaux doivent permettre, quant à eux, la modélisation du comportement.

La construction de SRM s'est faite en deux temps : Une première étape qui a consisté à spécifier tous les concepts utiles pour modéliser notre domaine d'intérêt, à savoir dans ce cas les plates-formes logicielles d'exécution temps-réel. Le résultat de cette étape s'est concrétiser par la définition d'un métamodèle (aussi souvent appelé dans ce cas, le modèle de domaine) ; la seconde étape a alors consisté à concevoir un support à la spécification du langage précédemment spécifié sous la forme d'une extension légère de UML, c.a.d. un profil UML (Selic, 2007).

## 4.1. Le modèle de domaine de SRM

SRM est basé sur la notion de ressource qui offre des services et peut posséder des propriétés (figure 4). Ces ressources et ces services peuvent être annotés par des propriétés non-fonctionnelles, qui permettent de quantifier et de qualifier les performances de la plate-forme d'exécution considérée. Un chapitre spécifique de

---

[1] www.promarte.org
[2] étiquette

MARTE est consacré à la description de ces propriétés non-fonctionnelles (voir chapitre 8 de (OMG, 2007)).

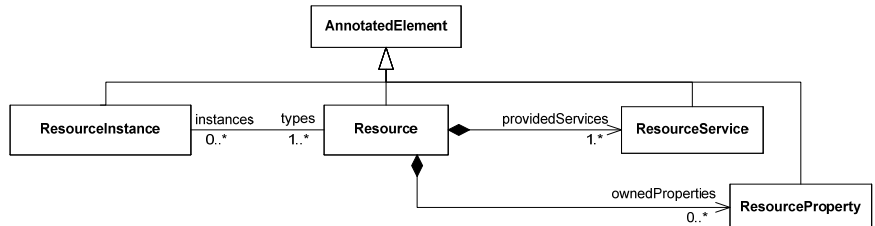

**Figure 4.** *Extrait du modèle de domaine de GRM : Les ressources et services*

Le concept de ressource est ensuite spécialisé pour le domaine du logiciel (figure 5). Les ressources fournissent donc un ensemble de services qui jouent des rôles spécifiques comme par exemple la création et la destruction de la ressource.

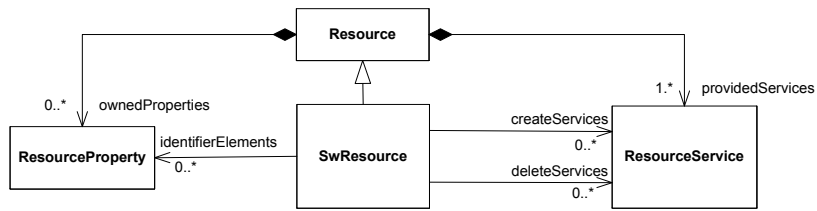

**Figure 5.** *Extrait du modèle de domaine de SRM : Les ressources logicielles*

Dans (Thomas, 2005), nous avons mené une étude détaillée des ressources et services offerts par les systèmes d'exploitation multitâches très largement utilisés dans l'industrie ainsi que par des standards d'interface de programmation tels que POSIX, OSEK/VDX et ARINC-653. Un aperçu des ressources logicielles représentatives des plates-formes d'exécution de ce domaine est présenté dans le tableau 1. De cette étude, il a été possible d'identifier les trois grandes familles de concepts suivantes :

- Les ressources concurrente, elles offrent des contextes d'exécution concurrents (par ex., tâche et interruption) ;
- les ressources d'interaction, elles proposent des mécanismes pour communiquer et se synchroniser des contextes d'exécution concurrents (par ex., sémaphore et boîtes aux lettres) ;
- les ressources de gestion, elles permettent de gérer les autres ressources logicielles et matérielles (par ex., les ordonnanceurs et les pilotes).

| Resource | Sémantique |
|---|---|
| **Resource concurrente** | |
| SchedulableResource | Encapsule une séquence d'action pour des exécutions concurrentes dont l'ordre et le temps d'exécution sont gérés par un ordonnanceur |
| MemoryPartition | Des espaces de nommage. |
| InterruptResource | Un contexte pour l'exécution des routines connectées à des signaux émis par l'environnement ou le système lui-même de manière asynchrone. |
| Alarm | Un contexte d'exécution pour une routine connecté à un réveil. |
| **Resource d'interaction** | |
| MessageComResource | Ressource utilisée pour échanger des messages (c.a.d. structure de données) |
| SharedDataResource | Ressource utilisée pour partager un espace mémoire |
| NotificationResource | Ressource utilisée pour notifier des occurrences d'événement |
| MutualExclusionResource | Ressource utilisée pour protéger des accès simultanés à un même espace de mémoire. |
| **Resource de gestion** | |
| MemoryBroker | Gestion de l'allocation et de partage de la mémoire |
| Scheduler | Gestion du séquencement des exécutions des ressources ordonnançables |
| DeviceBroker | Gestion des périphériques externes |

**Tableau 1.** *Types de Ressource du profil MARTE::SRM*

Des études similaires ont été menées sur la caractérisation des services et des propriétés. Elles ont abouti à une liste de services et propriétés représentatives du domaine métiers. La figure 6 est extraite de MARTE et présente l'aspect du modèle de domaine de SRM dédié à la description des ressources d'exécution concurrentes. Typiquement, une ressource d'exécution concurrente peut être périodique, apériodique ou sporadique (modélisée par la propriété nommée *type* de *ConcurrentResource* et typée par l'énumération *OccurenceKind*). Une ressource concurrente offre des services pour activer, suspendre, reprendre et terminer une exécution. Elle peut également être caractérisée par des propriétés permettant d'exprimer la priorité d'exécution, la période (pour des exécutions périodiques) ou enfin la taille de pile.

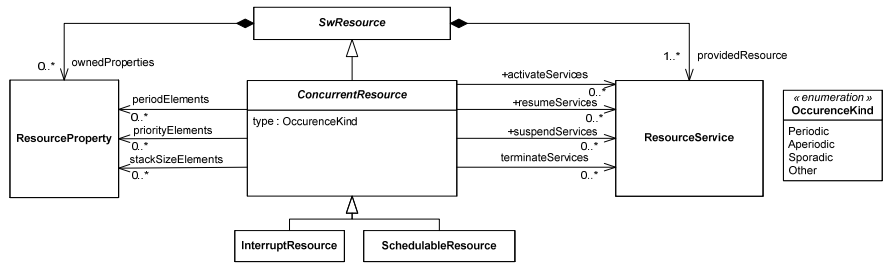

**Figure 6.** *Extrait du modèle de domaine de MARTE:: SRM*

## 4.2. Principes d'implantation du profil MARTE::SRM

En raison de la limitation en taille de l'article, nous ne présentons pas ici tous les détails du profil UML SRM (le lecteur est invité à se reporter à la norme elle-même pour cela). Un stéréotype pour chaque concept du tableau 1 a été créé. Ces stéréotypes étendent la métaclasse *Classifier* d'UML. Ils seront donc utilisés pour spécifier la sémantique des métaclasses filles de *Classifier* comme par exemple les classes, les composants et les collaborations.

Un exemple d'extension d'UML est présenté en figure 5. Il s'agit des stéréotypes permettant la description des ressources d'exécution concurrentes. Dans cette implantation du modèle de domaine, les propriétés *resourceProperty* et *resourceService* n'ont pas été transformées en stéréotype car les concepts natifs d'UML *TypedElement* et *BehavioralFeature* étaient suffisant. Une description détaillée est disponible dans la spécification du profile MARTE (voir chapitre 14 de (OMG, 2007)). Dans cet exemple, un élément conforme à un descendant de la métaclasse *Classifier* peut être spécifié comme étant un contexte d'exécution géré par un ordonnanceur (c.a.d. *SwSchedulableResource*) ou bien comme étant un contexte d'exécution (c.a.d. *InterruptResource*) dans lequel est exécuté un ensemble d'actions suite à un événement asynchrone du système (c.a.d. un signal).

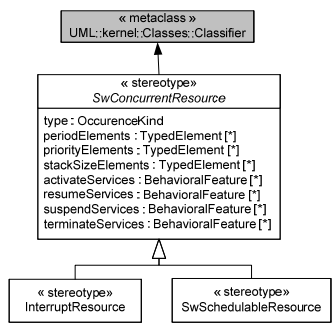

**Figure 7.** *Extrait du profile MARTE::SRM*

### 4.3. Un exemple d'utilisation du profil MARTE::SRM

L'exemple décrit en figure 8 illustre l'utilisation du profile MARTE::SRM pour la modélisation du concept de tâche basique (*BasicTask*) offert par la plate-forme d'exécution OSEK/VDX (OSEK/VDX Group, 2005). Une *BasicTask* offre un contexte pour exécuter de manière apériodique un traitement spécifié par l'utilisateur sous forme d'une routine (c.a.d. une fonction). L'ordre d'exécution et le temps alloué à la tâche pour exécuter cette routine sont gérés par un ordonnanceur logiciel. Le profile SRM est donc appliqué sur le paquetage nommé OSEK/VDX afin de caractériser les *BasicTask* comme étant une ressource ordonnançable, c.a.d. au sens de SRM une *SwSchedulableResource*.

L'API de OSEK/VDX définit ainsi pour le concept de tâche basique un ensemble normalisé de services et de propriétés statiques. Par exemple, une tâche basique propose les services *ActivateTask* et *TerminateTask* permettent respectivement d'activer et de terminer l'exécution d'une tâche. De même, les propriétés *Priority* et *StackSize* permettent à l'utilisateur de préciser la priorité d'ordonnancement et la taille de la pile allouée à une tâche basique.

Du point de vue du langage UML, les tâches basiques sont des classes stéréotypées *SwSchedulableResource*. Les sémantiques des services et des propriétés de tâches basiques sont précisées en renseignant les propriétés du stéréotype. Par exemple, l'attribut nommé *priority* permet d'exprimer la priorité de l'entité concurrente. Cet attribut est donc référencé par la propriété *priorityElements* du stéréotype *swSchedulableResource*. Ces sémantiques sont alors connues aussi bien des utilisateurs des plates-formes, que d'outils devant manipuler ces modèles de plate-forme. Il est d'ailleurs possible d'exprimer plusieurs éléments qui possèdent la même sémantique et plusieurs sémantiques pour le même élément. Par exemple, le service *ChainTask* permet à la fois de terminer une exécution et d'activer une autre exécution. Il est donc référencé par les deux propriétés *activateServices* et *terminateServices*.

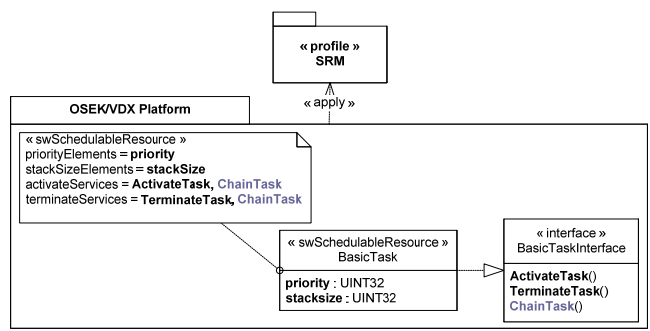

**Figure 8.** *Extrait du modèle de la plate-forme d'OSEK/VDX fait avec MARTE::SRM*

**4.4. Résultats**

Cette étude sur la caractérisation d'un métamodèle de plate-forme d'exécution logicielle pour le domaine du temps réel embarqué, nous a donc conduit à décrire le modèle du langage SRM. Il permet de décrire les ressources et services offerts par les plates-formes logicielles d'exécution multitâches. Nous avons testé dans (OMG, 2007) son application pour la description de deux librairies OSEK/VDX et ARINC-653. Ces librairies ont tout t'abord été modélisées en UML. Nous nous sommes concentrés sur la modélisation des ressources et des services par l'intermédiaire de diagrammes de classes. Ensuite nous avons précisé la sémantique de ces classes. Le profile SRM permet de spécifier la sémantique de 70 % des ressources offertes par la plate-forme OSEK/VDX et de 80% des ressources offertes par la plate-forme ARINC-653. Le profile SRM n'est donc pas exhaustif. En effet, des concepts liés à la modélisation des mécanismes de temps et à la gestion des modes de fonctionnement n'ont pour le moment pas été pris en compte dans le profile SRM. Les mécanismes de temps sont par exemple gérés par un autre profile de MARTE (voir chapitre 9 de (OMG, 2007)) qui n'a pas été appliqué dans cette étude. Il est donc nécessaire d'utiliser plusieurs métamodèle pour décrire une plate-forme d'exécution, chacun de ces métamodèles étant dédié à une considération métier donnée (par ex., la tolérance aux fautes pour les modes de fonctionnement).

**5. Intérêts d'une modélisation explicite de la plate-forme**

L'un des apports attendus d'une description explicite de la plate-forme en utilisant un métamodèle de plate-forme est la description de transformations indépendantes des plates-formes modélisées. Nous avons donc expérimenté une transformation permettant d'aider au portage sur d'autres plates-formes d'exécution d'une application multitâche d'ores et déjà dédiée à une plate-forme. Nous nous somme restreins à des modèles structurels.

La figure 9 montre un exemple très simplifié d'une application de robotique. Cette application a tout d'abord été décrite indépendamment d'une plate-forme d'exécution (paquetage *RobotController*). Elle a ensuite été déployée manuellement sur la plate-forme d'exécution OSEK/VDX. Pour cela l'application importe la description de la plate-forme (paquetage OSEK/VDX) pour pouvoir instancier une *BasicTask* de priorité *10*, nommé *t1* (paquetage SpecificOSEK/VDXApplication), et préciser que le code exécuté par cette tâche est la routine *trajectoryControl* de l'instance *C1*.

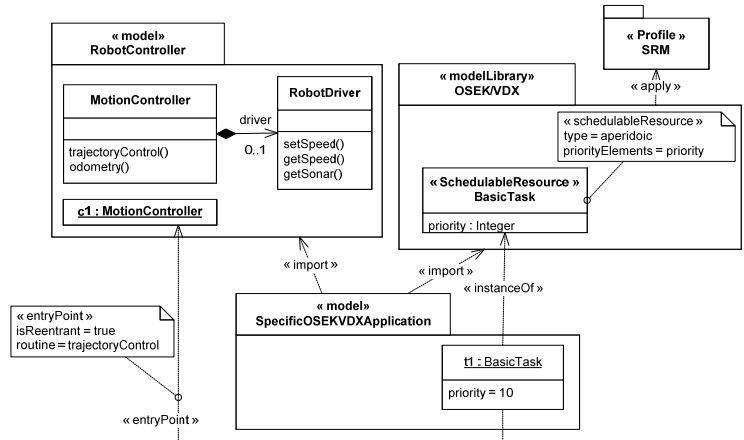

**Figure 9.** *Description d'une application dédiée à la plate-forme OSEK/VDX*

Dans cet exemple simple, nous souhaitons désormais aider au portage de cette application sur différentes plates-formes d'exécution. Nous souhaitons donc automatiser par transformation de modèle une partie du portage. Nous considérons une nouvelle plate-forme d'exécution : la plate-forme ARINC-653. La description de cette transformation consiste donc à transformer un modèle spécifié pour OSEK/VDX en un modèle spécifié pour ARINC-653. Une partie du modèle de cette plate-forme ARINC-653 est présentée en haut à droite de la figure 11. De manière générale, toutes les spécifications d'instances dont le type est stéréotypé *SwSchedulableResource* dans la plate-forme source (c.a.d. OSEK/VDX ici) doivent être transformées en des spécifications d'instances dont le type est stéréotypé *SwSchedulableResource* dans la plate-forme cible (c.a.d. ARINC-653). Un algorithme simplifié de la transformation est détaillé en figure 10.

**Pour** chaque spécification d'une instance dont le type est stéréotypé dans la plate-forme source **faire**

    **S'il** existe un type stéréotypé de la même façon dans la plate-forme cible **alors**

        **Si** les propriétés booléennes et énumérées des stéréotypes sont identiques **alors**

            Instancier ce type

            Renseigner le nom de l'instance cible par le nom de l'instance source

            Transformer les dépendances de l'instance cible par les dépendances de l'instance source

            **Pour** chaque propriété du type source référencée par une étiquette du stéréotype **faire**

                **Si** cette étiquette est renseignée dans le stéréotype cible **alors**

                    Convertir la valeur de cette propriété selon le type de la propriété référencée dans la plate-forme cible

                    Affecter cette valeur à la propriété référencée dans la plate-forme cible

                **Fin si**

            **Fin faire**

        **Fin si**

    **Fin si**

**Fin faire**

**Figure 10.** *Extrait de l'algorithme de transformation générique*

Comme dans la plate-forme OSEK/VDX, le concept de *BasicTask* est identifié comme une tâche ordonnançable apériodique, l'algorithme va tout d'abord recherché dans la plate-forme cible une ressource d'exécution apériodique. L'entité nommée *Process* est stéréotypée de cette façon dans la plate-forme ARINC-653. Par conséquent, pour chaque spécification dont le type est *BasicTask*, l'algorithme va définir dans le modèle cible une spécification d'instance dont le type est *Process*. Dans notre exemple, une spécification d'instance nommée *t1* typée *Process* est donc créée. Il recherche ensuite des concordances entre les propriétés des stéréotypes. La propriété *priorityElements* est renseignée dans les descriptions des plates-formes sources et cibles. Les attributs *priority* et *prio* possèdent donc la même sémantique. Ils sont donc tissés et peuvent par conséquent être transformés. L'algorithme va donc renseigner la priorité du *Process t1* à la valeur *10*.

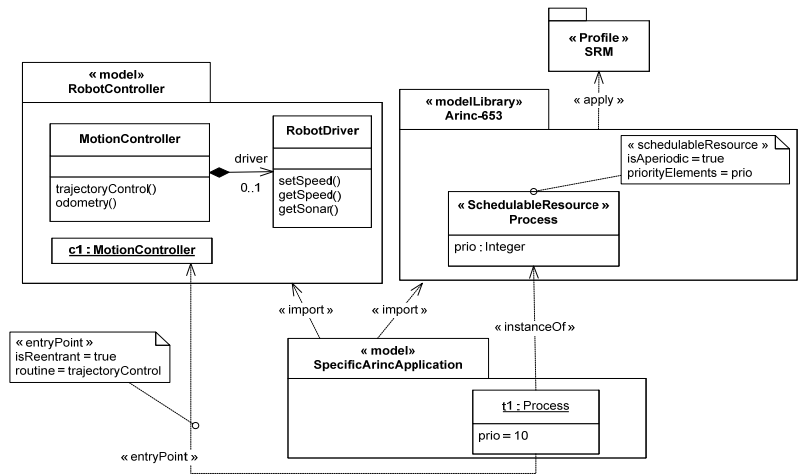

**Figure 11.** *Description d'une application dédiée à la plate-forme ARINC-653*

Cet algorithme a été implanté avec le langage ATL (Jouault, 2006) qui est un langage dédié à l'écriture de règles de transformations de modèle. L'algorithme est cependant limitée à un élément source correspondant à un élément cible. De plus, il reste limité aux ressources pour lesquelles la sémantique des concepts est capturée par une seule étiquette de stéréotype. Il ne traite pas plusieurs éléments ayant la même sémantique (c.a.d. plusieurs éléments renseignant la même propriété du stéréotype). Enfin, cet algorithme reste naïf. Il assure que deux éléments ayant la même sémantique sont tissés. Par contre, il n'assure pas que les valeurs associées à ces propriétés peuvent aussi être tissées de la même façon. Dans l'exemple précédent rien ne précise que la valeur de la priorité à *10* dans la plate-forme ARINC-653 possède la même sémantique que la valeur *10* dans la plate-forme

OSEK/VDX. La priorité *10* est une priorité élevée dans la plate-forme OSEK/VDX, ce qui n'est pas forcément le cas dans la plate-forme ARINC-653.

Dans cet exemple, il manque donc des informations dans la modélisation de la plate-forme pour caractériser les types des données. Un seul métamodèle métier (par ex. SRM pour le temps réel) pour décrire les plates-formes d'exécution n'est donc pas suffisant. Plusieurs métamodèles de plate-forme utilisés conjointement sont nécessaires pour décrire une même plate-forme d'exécution. R. Marvie propose dans (Marvie *et al.*, 2006) des opérateurs tel que la fusion ou l'agrégation. Ces opérateurs peuvent permettre de conserver des métamodèles de plates-formes dédiés à un domaine métier donné (par ex., le temps réel et les types de données).

Cette première expérience, nous a tout de même confirmé l'intérêt d'un modèle de plate-forme explicite. Une description explicite de la plate-forme nous permet de séparer efficacement les préoccupations métiers dans l'écriture des transformations. Par exemple, il n'y a aucunes connaissances métiers liées au multitâche dans l'algorithme proposé. Celles-ci sont décrites au niveau des modèles. Un métamodèle de plate-forme nous permet donc bien de séparer les préoccupations liées d'une part au langage de modélisation et d'autre part aux caractéristiques et aux règles métiers des plates-formes considérées lors de la description de transformations.

## 6. Un métamodèle de plate-forme d'exécution : un mythe ou une réalité ?

D'autres travaux se sont intéressés à décrire des plates-formes d'exécution qu'elles soient logicielles ou matérielles (Kukkala *et al.*, 2005), (Taha *et .al*, 2007) Même si ces travaux utilisent une description de la plate-forme pour différents objectifs (c.a.d. analyse de performance, déploiement, simulation matérielle), ils nécessitent un modèle explicite de la plate-forme. Ils définissent donc un métamodèle pour décrire des plates-formes. Nous avons reproduit dans les figures 12 et 13 l'approche commune suivie dans la littérature. Nous la comparons à l'approche utilisée pour la conception de SRM. La partie gauche des figures correspond au motif de modélisation que nous avons utilisé pour SRM dans MARTE. La partie droite correspond au motif présent dans la littérature.

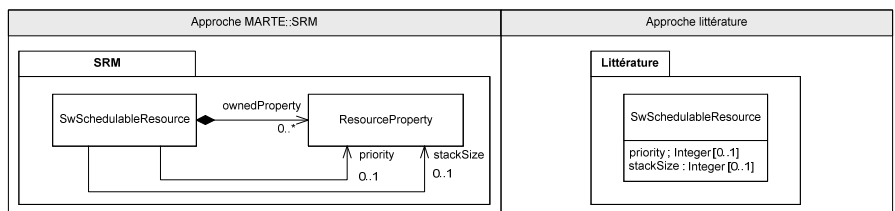

**Figure 12.** *Modèles de domaine pour la modélisation des plates-formes d'exécution*

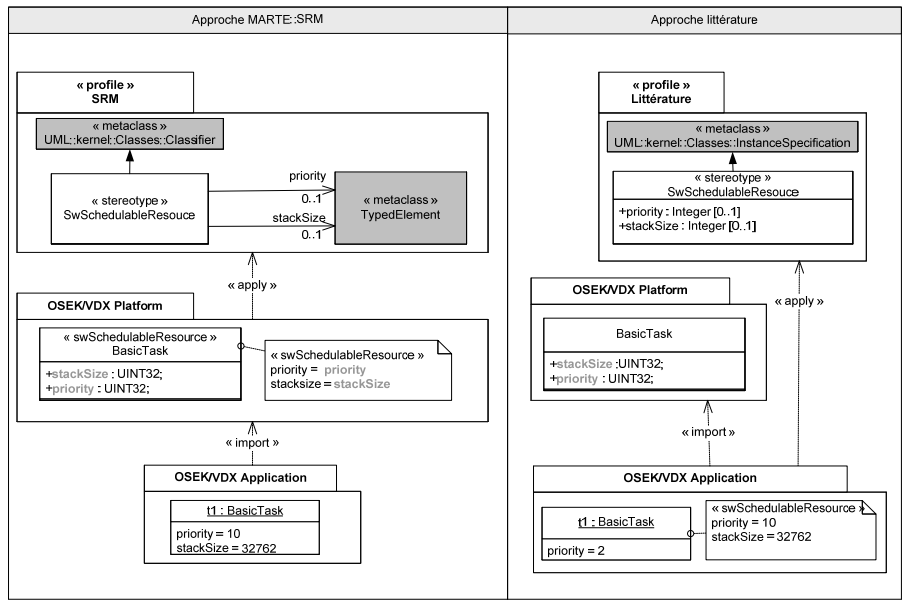

**Figure 13**. *Profiles UML pour la modélisation des plates-formes d'exécution*

Il apparaît sur cette figure que les deux approches sont similaires. Néanmoins, dans l'approche de la littérature, le métamodèle n'est pas un langage pour décrire des plates-formes d'exécution. En effet, les propriétés des ressources sont d'ores et déjà typées dans le métamodèle *Littérature*. Dans l'approche de la littérature la *priorité* d'une tâche est un entier. Dans le profile résultant les stéréotypes étendent, entre autre chose, la sémantique des *InstanceSpecification* d'UML. Ces stéréotypes sont donc applicables sur les spécifications d'instance. Ils permettent alors de décrire les caractéristiques de l'application mais pas celles de la plate-forme d'exécution. Par exemple, la priorité de la tâche *t1* est précisée par les propriétés du stéréotype et non par les propriétés de son type *BasicTask*. En termes de modélisation, c'est l'*InstanceSpecification* qui possède la priorité et non l'entité *t1*. Cela peut donc conduire à des incohérences dans le modèle comme l'illustre la figure 13. Ainsi, le type de l'attribut *priority* de la classe *BasicTask* n'est pas le même que celui de la propriété *priority* du profile *Littérature*. De même, il n'existe pas de concordance entre les valeurs renseignées pour ces deux informations qui sémantiquement expriment la même chose : la priorité de la tâche ; mais qui peuvent prendre des valeurs différentes. Le profile *Littérature* est une nouvelle API permettant de décrire des applications multitâches.

Au sens des travaux de J.P Almeida (Almeida, 2006), le métamodèle est donc une plate-forme abstraite. Sa description correspond à celle d'une plate-forme fictive, idéale, permettant de décrire des applications. Pourtant, un modèle de langage dédié à la description des plates-formes d'exécution n'a pas pour objectif de fournir une nouvelle API. Ce doit être un langage permettant la description de cette API. En utilisant le profile SRM, ce sont bien les ressources qui sont décrites et non leur utilisation pour une application donnée. D'ailleurs, le métamodèle SRM peut être utilisé pour décrire le métamodèle *Littérature*. Le métamodèle SRM permet la description de la plate-forme abstraite nommée *Littérature*. Par contre, le métamodèle *Littérature* ne permet par la description du métamodèle SRM. Ce n'est donc pas un métamodèle de plate-forme mais un modèle de plate-forme.

L'un des apports attendus d'une description explicite de la plate-forme en utilisant un métamodèle de plate-forme est la description de transformations indépendantes des plates-formes modélisées. L'approche de la littérature ne permet pas de le faire car les transformations sont dédiées à cette plate-forme abstraite *Littérature*. Il n'y a plus séparation des préoccupations dans l'écriture des transformations.

Le modèle de plate-forme est donc bien une réalité si l'on considère que les propriétés et les services offerts par les ressources de la plate-forme sont des rôles joués. Cela permet alors de structurer efficacement le métamodèle. Une synthèse de cette structure est présentée en figure 14.

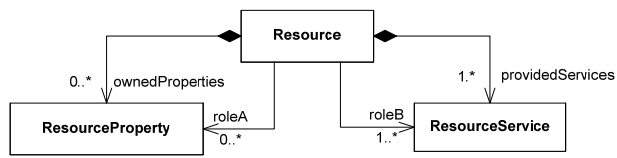

**Figure 14.** *Structure des métamodèles de plates-formes d'exécution*

## 7. Conclusion

Dans cet article, nous avons contribué à la modélisation explicite des plates-formes d'exécution. Pour cela nous avons tout d'abord définit ce que la communauté de l'IDM considérait comme étant une plate-forme d'exécution. Au sens de l'IDM, le concept de plate-forme d'exécution est un serveur de ressources et de services pour d'autres systèmes appelés applications. A partir des travaux de la communauté, nous avons ensuite caractérisé les grands axes nécessaires à la conception d'un métamodèle de plate-forme : la caractérisation des ressources, la caractérisation des services, la caractérisation des règles d'utilisation et la caractérisation du comportement. Nous nous sommes concentrés sur la caractérisation des deux premiers axes par la contribution à un métamodèle des ressources et services des plates-formes d'exécution multitâches. Ce métamodèle est implanté dans un profil UML, nommé Software Resource Modeling (SRM). Même, s'il est intégré au nouveau profil UML standard, de l'OMG, dédié à la modélisation et à l'analyse des systèmes embarqués temps réel (MARTE), ce profil n'est pas exhaustif. Seule l'expérimentation peut nous permettre de le valider. Nous l'avons donc testé sur plusieurs plates-formes d'exécution. Le niveau d'abstraction de SRM est adapté pour la modélisation des plates-formes que nous ciblons. Néanmoins, il ne permet pas d'exprimer toutes les caractéristiques de ces plates-formes. Il doit donc être utilisé conjointement avec d'autres métamodèles de plate-forme métier (par ex., la sureté de fonctionnement et le typage des données). Des opérateurs tels que la fusion ou l'agrégation de ces métamodèles de plates-formes sont à étudier.

Nous avons dans une deuxième partie expérimentée un premier algorithme de transformation manipulant des modèles de plate-forme explicites. Cet algorithme est limité aux transformations structurelles de ressource. Il nous a permis de montrer l'intérêt d'une modélisation explicite de la plate-forme en soulignant le caractère générique de la transformation et en révélant la séparation des préoccupations lors de la description de cette transformation. Même s'il reste néanmoins énormément de configuration à traiter et à valider, il nous faudra par la suite considérer dans de telles transformations un ensemble de métamodèle de plate-forme.

SRM a été conçue suivant un motif particulier lié à la modélisation de plate-forme. Nous avons discuté différentes approches pour la description d'un tel métamodèle de plate-forme d'exécution. Nous en avons conclu que le modèle de plate-forme est bien une réalité si l'on considère que les propriétés et les services offerts par les ressources de la plate-forme sont des rôles joués. Cette structure peut être utilisée pour décrire de nouveaux métamodèles de plate-forme. Que se soit dans la technologie des profiles UML ou de langages dédiés (Domain Specific Language (DSL)), les concepts de ce motif peuvent être utilisés et spécialisés pour un domaine métier donné.

Nos efforts se portent désormais sur l'intégration d'un ensemble de métamodèles de plates-formes d'exécution dans le cycle de développement d'application temps réel embarqués. Nous n'avons pas jusque là validé l'intérêt du schéma ressource-service annoté par des propriétés permettant de caractériser les performances et les qualités de service. Nous l'avons pour le moment accepté comme point d'entré pour nos métamodèles de plate-forme d'exécution. Enfin, nous devons aussi orienter nos recherches sur les liens qui peuvent être mis en œuvre pour lier l'application et la plate-forme et évaluer leurs implications sur des transformations génériques permettant de passer d'une plate-forme à une autre.

## 8. Bibliographie